\documentclass[pre,aps,twocolumn,showpacs,10pt]{revtex4-1}
\usepackage{amsmath,amssymb,graphicx}
\usepackage{epsfig}
\usepackage{textcomp}
\usepackage{color}
\usepackage{epstopdf}
\usepackage{array}
\usepackage[normalem]{ulem}
\usepackage{amssymb}
\usepackage{amsmath}
\usepackage{graphicx}

\usepackage{mathrsfs}
\usepackage[utf8]{inputenc}
\usepackage[english]{babel}
\usepackage{amsthm}

\begin{document}
\title{Population Extinction under Bursty Reproduction in a Time Modulated Environment}
\author{Ohad Vilk and Michael Assaf}
\email{michael.assaf@mail.huji.ac.il}

\affiliation{Racah Institute of Physics, Hebrew University of Jerusalem, Jerusalem 91904, Israel}

    \begin{abstract}
    In recent years non-demographic variability has been shown to greatly affect dynamics of stochastic populations. For example, non-demographic noise in the form of a bursty reproduction process with an a-priori unknown burst size, or environmental variability in the form of time-varying reaction rates, have been separately found to dramatically impact the extinction risk of isolated populations. In this work we investigate the extinction risk of an isolated population under the \textit{combined} influence of these two types of non-demographic variation. Using the so-called momentum-space WKB approach we arrive at a set of time-dependent Hamilton equations. In order to account for the explicit time dependence, we find the instanton of the time-perturbed Hamiltonian numerically, where analytical expressions are presented in particular limits using various perturbation techniques. We focus on two classes of time-varying environments: periodically-varying rates corresponding to seasonal effects, and a sudden decrease in the birth rate corresponding to a catastrophe. All our theoretical results are tested against numerical Monte Carlo simulations with time-dependent rates and also against a numerical solution of the corresponding time-dependent Hamilton equations.

    \end{abstract}
     \maketitle

\section{Introduction}
Stochastic processes that result in the extinction of a stochastic population after maintaining a long-lived state, affect a wide range of biological populations, and have attracted much interest over the past decades. Manifestations of such stochastic processes range from population biology, epidemiology, cell biochemistry, virology, gene regulation and conservational ecology, see \textit{e.g.} \cite{bartlett1978introduction,horsthemke1984noise,gardiner1985handbook,van1992stochastic,parzen1999stochastic,
assaf2008noise,schwartz2009predicting,pearson2011stochastic}.

If the population is isolated then there is always an absorbing state at zero. That is, extinction can occur due to a rare sequence of death events owing to \textit{demographic noise}, which stems from the stochastic nature of the reactions and discreteness of individuals. While most previous studies of population extinction have focused on this type of noise, see e.g. Refs.~\cite{Dykman1994,nisbet2003modelling,elgart2004rare,doering2005extinction,assaf2006spectral,assaf2007spectral,kessler2007extinction,
assaf2008population,assaf2009population,assaf2010extinction,assaf2010large,meerson2013immigration,be2015colonization}, non-demographic variability (see \textit{e.g.} Refs.~\cite{hanggi1995colored,elowitz2002stochastic}) may dramatically influence the extinction risk of a population subject to demographic noise \cite{leigh1981average,lande1988genetics,lande1993risks,lande1998extinction,kamenev2008colored,assaf2013extrinsic,
levine2013impact,Bacaeer2015,be2016effect,be2016rare,assaf2017wkb,billings2017seasonal,be2017enhancing}. In general, non-demographic noise originates from the variability across individuals as well as from environmental variations, and can give rise to time-varying reaction rates. These variations, however, are not necessarily stochastic and may be caused by deterministic factors such as seasonal shifts in temperature or humidity, competition, breeding sites, or forage availability, see e.g. \cite{martin2013reciprocal,gonzalez2013variability, crespi2013life}. Notably, while these factors result in time-periodic reaction rates~\cite{lande1988genetics,assaf2008population,Bacaeer2015,billings2017seasonal}, a population can also experience a sudden drastic drop in the birth rate due to a drastic deterioration of environmental conditions~\cite{assaf2009population}.

In addition to varying the reaction rates, non-demographic noise can also influence the reaction \textit{step size}. Here, for example, instead of having a single birth event with a fixed number of products $A\to 2A$, non-demographic noise can give rise to a bursty reproduction process $A\to A+kA$, where $k=0,1,2...$ is a random non-negative integer that is drawn from a given step-size distribution. This type of uncertainty, or noise, appears in a wide variety of scientific areas including population biology and ecology~\cite{goel2016stochastic}, viral dynamics~\cite{pearson2011stochastic}, and cell biology~\cite{paulsson2000random,shahrezaei2008analytical}. Importantly, such reaction step-size noise contributes to the variability in the ecological traits of a population and can strongly affect the extinction probability of a population~\cite{gonzalez2013variability,be2016effect,be2017enhancing}.

In previous works, the extinction risk of a population has been studied separately under the influence of deterministically time-varying rates~\cite{assaf2008population,assaf2009population,Bacaeer2015,billings2017seasonal}, and reaction step-size noise~\cite{be2016rare,be2016effect,be2017enhancing}. In reality, however, these effects should both be taken into account. For example, seasonal fluctuations of temperature can cause a time modulation in the reaction rates, while variations in the offspring number per birth event (which also depends on the seasonal variability) cause an uncertainty in the reaction step size. As a result, in this paper we study the combined effect of time-varying reaction rates and uncertainty in the reaction step size on the extinction risk of a population, thus generalizing previous works in this field~\cite{assaf2008population,assaf2009population,Bacaeer2015,billings2017seasonal}. For concreteness we employ the generalized version of the Verhulst logistic model with bursty reproduction~\cite{be2016effect,be2017enhancing}, with time-dependent rates, and calculate, using the so-called momentum-space WKB approach~\cite{elgart2004rare,assaf2006spectral,assaf2007spectral,assaf2010large},  the mean time to extinction (MTE) for generic step-size distributions (SSDs).

The paper is organized as follows. In Sec. \ref{Background}, we employ the generating function formalism in order to transform the master equation into a partial differential equation for the probability generating function. Then we apply the eikonal method to this equation, which yields in the leading order a Hamilton-Jacobi equation with an effective (explicitly) time-dependent Hamiltonian. The latter also accounts for the uncertainty in the reaction step size due to the bursty reproduction. To this end, we analyze the corresponding Hamiltonian in various limits. In Sec. \ref{Sinusoidal} we follow Assaf et al. \cite{assaf2008population}  and apply three perturbation techniques, in three different regimes, to a population with time-periodic rates. The first regime is when the modulation amplitude is small, and a linear theory (LT) with respect to the modulation amplitude can be applied (Sec. \ref{LT_section}). In the second regime, in the limit of high modulation frequency, we employ a formalism in the spirit of the Kapitsa method \cite{landau1976mechanics} (Sec. \ref{Kapitsa}), while in the third low-modulation frequency regime, we employ an adiabatic theory (Sec. \ref{Adiabatic}). Furthermore, in Sec. \ref{Catastrophe} we consider a different time dependence of the reaction rates in the form of a finite and predetermined drop in birth rates, and compute the corresponding increase in the extinction risk of the population \cite{assaf2009population}. We dedicate Sec. \ref{NumericalCalc} to a short description of the time-dependent Monte Carlo simulation that we have used as well as to describe the method by which we solve the explicitly time-dependent Hamilton equations using the shooting method. Finally, in Sec. \ref{summary} we discuss the interplay between the two forms of non-demographic variability that  we have considered.

\section{Master Equation, Probability Generating Function and the Unperturbed Action} \label{Background}
Our starting point is the generalized Verhulst model with bursty reproduction~\cite{be2016rare}. The microscopic dynamics of our system are given by the following birth-death reactions with the corresponding rates:
\begin{eqnarray} \label{eq1}
&&A\overset{\lambda_{n}}{\to}A+kA\;;\;\;\lambda_{n}=B(t) n D(k)/\langle k\rangle;\;\;k=0,1,2,...\;,\; \nonumber\\
&& A\overset{\mu_{n}}{\to}\emptyset\;;\hspace{12mm}\mu_{n}=n+B_{0}n^{2}/N.
\end{eqnarray}
Here $n$ is the population size and $N \gg 1$ is the typical population size in the long-lived metastable state prior to extinction, see below. Also, the burst size $k$ is a-priori unknown and is drawn from a normalized SSD, $D(k)$, with a mean value of $\langle k \rangle$ and standard deviation $\sigma$. In addition, the birth rate per capita satisfies $B(t) = B_0 g(t)$, where $g(t)$ is a known function of time and $B_0 > 1$  is the average reproduction rate per capita.
\begin{figure}[t]
	\includegraphics[width=0.85\linewidth]{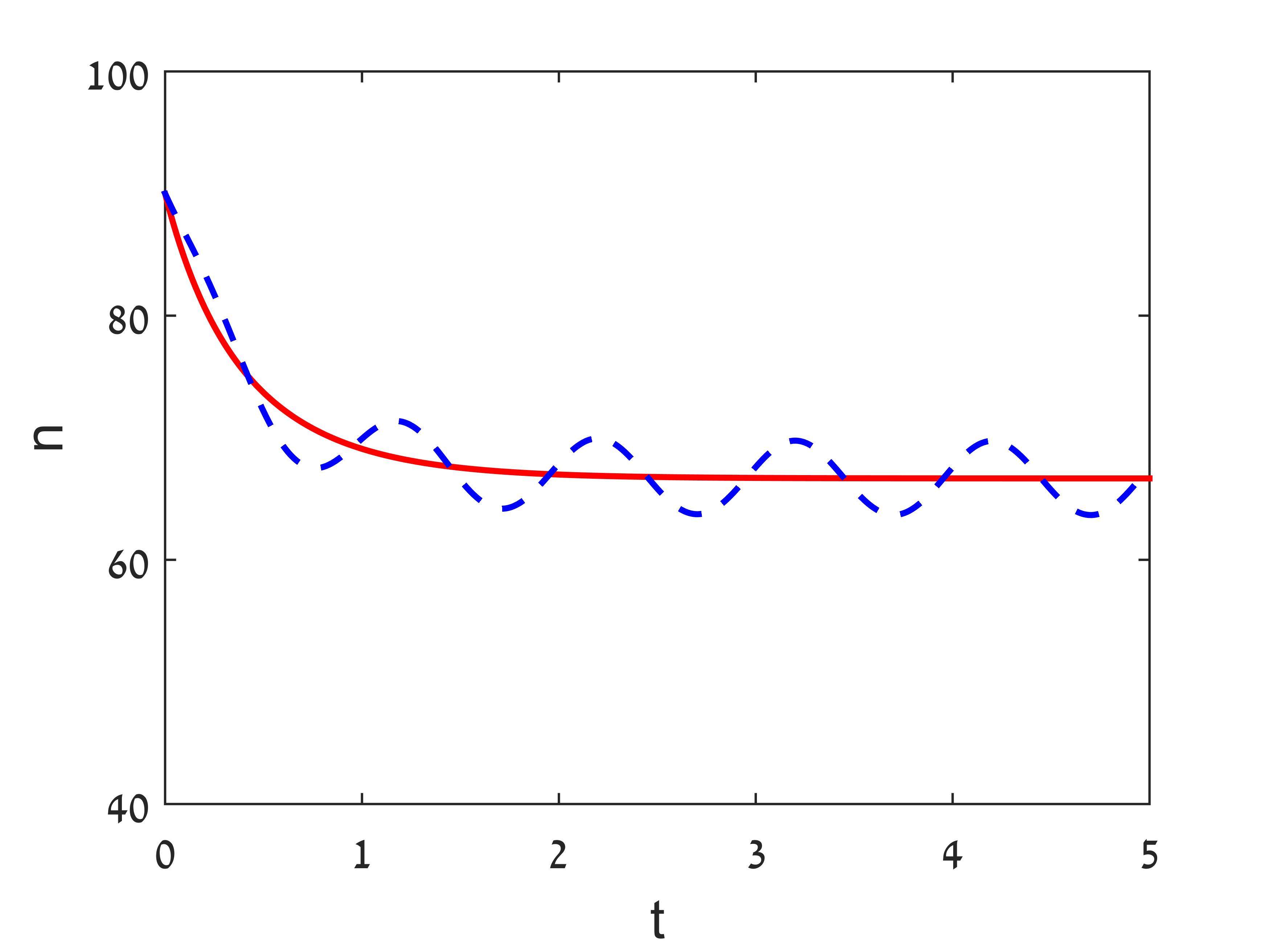}
	\caption{Numerical solution of the rate equation, see text: a comparison between the time-perturbed (solid line) and unperturbed (dashed) cases. Parameters are $B_0=3$ and $N=200$ for both cases, while the birth rate in the perturbed case is given by $B(t) = B_0 [1+\epsilon\cos(\omega t)]$ with $\epsilon=0.1$ and $\omega = 1$.}
	\label{fig:RateEquation}
\end{figure}

Using Eq.~(\ref{eq1}), the deterministic (mean-field) dynamics is governed by the following rate equation: $\dot{\bar{n}} = \bar{n}[B(t)-1-B_0\bar{n}/N]$. In the time-independent case, this equation has a stable fixed point at $n=N(B_0-1)/B_0$, and an unstable fixed point at $n=0$. Henceforth, we will assume that the typical population size at the stable fixed point satisfies $N\gg 1$. In Fig.~\ref{fig:RateEquation} we present the typical mean-field dynamics of $\bar{n}(t)$ for a periodic birth rate.

The rate equation ignores demographic fluctuations. To account for these, and to compute the MTE, we consider the master equation describing the time-evolution of the probability $\mathcal{P}_n(t)$ of having $n$ individuals at time $t$. Using Eq.~(\ref{eq1}) the master equation reads
\begin{eqnarray} \label{master}
&&\dot{\mathcal{P}}_n=\frac{B(t)}{\langle k\rangle}\left[\sum_{k=0}^{n-1}D(k)(n-k)\mathcal{P}_{n-k}-n\mathcal{P}_n\right]\nonumber\\ &&+(n+1)\mathcal{P}_{n+1}-n\mathcal{P}_n+\frac{B_0}{N}\left[(n+1)^2\mathcal{P}_{n+1}-n^2\mathcal{P}_n\right]\!.
\end{eqnarray}
Note that the rate equation described above can be obtained from this master equation by multiplying the latter by $n$, summing over all $n$'s, and using the definition $\bar{n}(t)=\sum_{n}n\mathcal{P}_n(t)$.

To treat master equation~(\ref{master}) we introduce the probability generating function~\cite{gardiner1985handbook} $G(\wp,t)=\sum_{n=0}^{\infty}\wp^n\mathcal{P}_n(t)$, with $\wp$ being an auxiliary variable. Note, that $\mathcal{P}_n(t)$ is given by the Taylor coefficients of $G(\wp, t)$ around $\wp=0$. Multiplying Eq.~(\ref{master}) by $\wp^n$ and summing over all $n$'s, we arrive at a partial differential equation for $G(\wp,t)$
\begin{equation}\label{Gfun}	
\frac{\partial G}{\partial t}=(\wp-1)\left\{ \left[B(t)\wp f(\wp)\!-\!1\!-\!\frac{B_{0}}{N}\right]\frac{\partial G}{\partial\wp}-\frac{B_{0}}{N}\wp\frac{\partial^{2}G}{\partial\wp^{2}}\right\}\!,
\end{equation}
where we have defined $f(\wp)=\sum_{k=0}^{\infty} D(k) (\wp^{k}-1)/[\left\langle k\right\rangle (\wp-1)]$, which is related to the probability generating function of the SSD. Assuming $N\gg 1$, employing the eikonal ansatz $G(\wp, t)\sim \exp\left[-N S(\wp , t )\right]$ in Eq. (\ref{Gfun}), where $S(\wp)$ is the action function~\cite{elgart2004rare}, and neglecting subleading-order terms with respect to $N$, we arrive at a Hamilton Jacobi equation
\begin{equation}
\frac{\partial S}{\partial t}=(\wp-1)\left\{ \left[B(t)\wp f(\wp)-1\right]\frac{\partial S}{\partial\wp}+B_{0}\wp\left(\frac{\partial S}{\partial\wp}\right)^{2}\right\}\!.
\end{equation}

Introducing a canonically conjugate coordinate $q=-\partial S/\partial\wp$, and shifting the momentum $p=\wp-1$, we arrive at the following one dimensional Hamiltonian flow, where $p$ plays the role of the momentum~\cite{elgart2004rare}:
\begin{equation} \label{Hamiltonian}
H(t)=pq\left[ B(t)(p+1)f(p+1)-1-B_{0}(p+1)q\right],
\end{equation}
and to remind the reader, $B(t)=B_0 g(t)$~\footnote{In the time-independent case, Eq.~(\ref{Hamiltonian}) coincides, up to a canonical transformation, with the Hamiltonian obtained by Be'er et el. in the real space coordinates, see Ref.~\cite{be2016rare}.}. The corresponding Hamilton equations are
\begin{multline} \label{Hamilton_q}
\dot{q}=q\left[ B(t)(2p+1)f(p+1)-1-B_{0}(2p+1)q\right. \\ \left.+Bp(p+1)f^{\prime}(p+1)\right]
\end{multline}
\begin{equation} \label{Hamilton_p}
\dot{p}=-p\left[ B(t)(p+1)f(p+1)-1-B_{0}(p+1)2q\right].
\end{equation}

When the rates are time independent, Hamiltonian~(\ref{Hamiltonian}) is conserved and the problem is integrable. In this case, the most probable path to extinction, often referred to as the optimal path to extinction or instanton~\cite{Dykman1994}, is a nontrivial zero-energy trajectory of (\ref{Hamiltonian}), and is given by:
\begin{equation} \label{Instanton}
q_0(p_0)=f(p_0+1)-\frac{1}{B_{0}(p_0+1)}.
\end{equation}
The corresponding action along the instanton satisfies
\begin{equation} \label{Szero}
S_{0}=-\int_{0}^{p_{f}}q_{0}(p)dp=\frac{1}{B_{0}}\ln(1+p_{f})-\int_{0}^{p_{f}}f(1+p)dp,
\end{equation}
where $p_f$ is the momentum associated with the fluctuational fixed point, $(q=0,p=p_f)$, which can be found by solving the transcendental equation $f(p_f+1)=1/(B_0(p_f+1))$. In the leading order, the MTE is given by $\tau_{ex}\sim \exp(N S_0)$~\cite{elgart2004rare}. Note, that having found $p_f$, the action $S_0$ can be evaluated by substituting the exact form of $f(p+1)$ into Eq. (\ref{Szero})~\cite{be2016rare}.

\section{Periodic Environment} \label{Sinusoidal}
Let us now assume that the time modulation is periodic, $g(t) = 1+\epsilon\cos(\omega t)$. The time-dependent Hamiltonian~(\ref{Hamiltonian}) is now given by
\begin{equation} \label{Hamiltonian_LT}
H(q,p,t)=H_{0}(q,p)+\epsilon H_{1}(q,p,t)
\end{equation}
where
\begin{equation} \label{H_0}
H_0(q,p)=p q\left[ B_0(p+1)f(p+1)-1-B_{0}(p+1)q\right]
\end{equation}
and
\begin{equation} \label{H_1}
H_{1}(q,p,t)=pqB_{0}(p+1)f(p+1)\cos(\omega t).
\end{equation}
To compute the MTE up to leading order we need to find the action along the perturbed instanton of the time-dependent Hamiltonian. Denoting the coordinates of the perturbed path as $q(t,t_{0})$ and $p(t,t_{0})$, a general expression for the action can be written as~\cite{assaf2008population}
\begin{multline} \label{S_exact}
S=\int_{-\infty}^{\infty}\left\{ p(t,t_{0})\dot{q}(t,t_{0})-H_{0}\left[q(t,t_{0}),p(t,t_{0})\right] \right.\\\left. -\epsilon H_{1}\left[q(t,t_{0}),p(t,t_{0}),t\right]\right\} dt,
\end{multline}
where $t_0$ is the phase element of the Poincare map which gives the minimal action \cite{dykman1997resonant,dykman2001activated,escudero2008persistence}. As mentioned above, this problem can be analytically solved only in specific limits. Following Ref.~\cite{assaf2008population} we henceforth apply three perturbation techniques in different parameter regimes.

\subsection{Linear Theory} \label{LT_section}

In this section we assume that the time perturbation is small, i.e., $\epsilon\ll 1 $. Let us define by  $q_{0}(t-t_{0})$ and $ p_{0}(t-t_{0})$ the coordinate and the momentum of the unperturbed zero-energy instanton evaluated at time $t-t_{0}$ \footnote{In the time-independent case, $q_0(t-t_0)$ and $p_0(t-t_0)$ can be found by solving Hamilton equations $\dot{q_0}=\partial_p H_0$, and $\dot{p_0}=-\partial_q H_0$. Here $t_0$ serves as an arbitrary time shift.}. For $\epsilon\ll 1$, it has been shown that the action can be approximated as \cite{dykman1997resonant,dykman2001activated,escudero2008persistence,assaf2008population}:
\begin{equation}
S(t_{0})\approx S_{0}+\Delta S(t_{0})
\end{equation}
where $S_0$ is the unperturbed action and
\begin{equation} \label{DeltaS}
\Delta S(t_{0})=-\epsilon\int_{-\infty}^{\infty}H_{1}\left[q_{0}(t-t_{0}),p_{0}(t-t_{0}),t\right]dt.
\end{equation}
To find the optimal correction, the action $S(t_0)$ must be \textit{minimized} with respect to $t_0$. Substituting Eq. (\ref{Instanton}) into Eq. (\ref{Hamilton_p}), we arrive at the following integral equation
\begin{equation} \label{problem}
t\left(p_0\right) = t_0 + \int^{p_0}\frac{dp_0}{p_0\left[ B_{0}(p_0+1)f(p_0+1)-1\right]}.
\end{equation}
Further, using Eqs. (\ref{Hamilton_p}) and (\ref{Instanton}) one finds $\dot{p}=B_{0}qp(p+1)$. Employing the latter and Eq. (\ref{H_1}), Eq. (\ref{DeltaS}) becomes
\begin{equation} \label{Delta_S_simple}
\Delta S(t_{0})=-\epsilon\int_{0}^{p_f}f(p_0+1)\cos(\omega t)dp_{0}
\end{equation}
where $p_f$ is the fluctuational momentum defined above, and $t$ is a function of $p_0$ as indicated by Eq. (\ref{problem}). Note, that a particular choice of the SSD, $D(k)$, determines the form of $f(p)$ in both Eq.~(\ref{problem}) and (\ref{Delta_S_simple}). Finally, to find the minimal action,  solution~(\ref{Delta_S_simple}) has to be minimized with respect to $t_0$. As a result, and as was previously shown by Dykman et al.~\cite{Dykman1994,dykman1997resonant,dykman2001activated}, we find that in the LT the modulation signal removes the degeneracy of the unperturbed instanton trajectories with respect to the arbitrary time shift $t_0$. It is thus possible to select the optimal instanton in relation to the modulation signal.

Having found $t_0$ for which the correction to the action is minimal, the MTE is given by
\begin{equation}\label{MTE}
\tau\sim e^{N(S_0+\Delta S)}.
\end{equation}
where $\Delta S=\min_{_{t_0}}[\Delta S(t_0)]$ is negative, and $\Delta S(t_0)$ is given by Eq.~(\ref{Delta_S_simple}). This indicates that the time modulation yields an exponential increase in the population's \textit{extinction risk}, by a factor of $e^{N|\Delta S|}$.

Before considering particular examples, let us discuss the validity of the LT. The condition for the general linear correction to hold is that $\epsilon\ll 1$ and
\begin{equation} \label{LT_condition}
S_{0}+\Delta S\gg 1/N.
\end{equation}
Strictly speaking we also need to separately demand that $|\Delta S|\gg 1/N$ for the eikonal approximation to hold \cite{assaf2008population}, but we have checked that the theory works well already when $N|\Delta S|\gtrsim {\cal O}(1)$, see below.

In the next two subsections we will find the explicit reduction of the action in two simple limits: the case of single-step reaction (SSR), $D(k)=\delta_{k,1}$, and for a general SSD close to the bifurcation limit.

\begin{figure}[t]
	\includegraphics[width=0.85\linewidth]{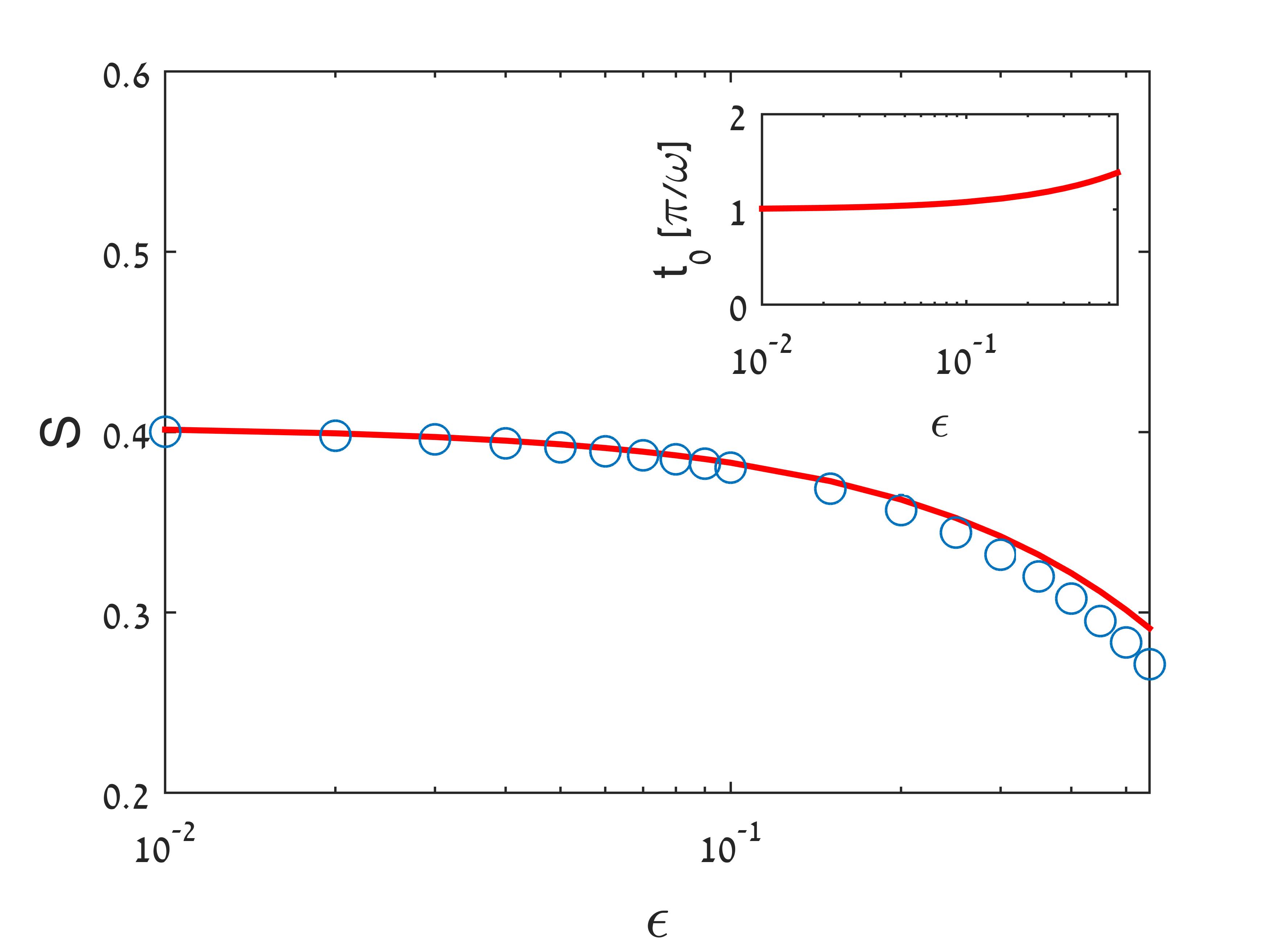}
	\caption{A comparison between the theoretical (solid line) and numerical (symbols) actions in the case of the SSR, in the LT regime, as a function of $\epsilon$. The numerical solution is obtained by numerically calculating the instanton trajectory of the perturbed Hamiltonian. The parameters are $B = 4$ and $\omega = 3$. Inset shows a numerical calculation of $t_0$ which minimizes the action, as a function of $\epsilon$, see Sec. \ref{NumericalCalc}. }
	\label{fig:ActionCalculationLT}
\end{figure}

\subsubsection{Linear theory - Single Step Reaction} \label{SSR_LT}

In the SSR case, $D(k)=\delta_{k,1}$, we substitute $f(p)=1$ into Eq.~(\ref{problem}) and find $t(p_0)$, which can then be plugged into Eq.~(\ref{Delta_S_simple}). After some algebra, $S(t_0)$ can be shown to satisfy the following integral
\begin{eqnarray} \label{SSRintegral}
\Delta S(t_{0})&=&\epsilon\frac{B_{0}-1}{B_{0}}\int_{-\infty}^{\infty}\frac{1}{1+e^{-x}}\frac{1}{1+e^{x}}\nonumber\\
&\times&\cos\left(\frac{\omega}{B_{0}-1}x+\omega t_{0}\right)dx,
\end{eqnarray}
which yields
\begin{equation}
\Delta S(t_{0})=\frac{\epsilon\pi\omega}{B_{0}}\cos(\omega t_{0})csch\left(\frac{\pi\omega}{B_{0}-1}\right).
\end{equation}
Minimizing the action with respect to $t_0$ we find that $t_0 = \pi/\omega$, which yields the correction to the action for the SSR case~\cite{Bacaeer2015}
\begin{equation} \label{S_SSR}
\Delta S_{SSR}=-\frac{\epsilon\pi\omega}{B_{0}}csch\left(\frac{\pi\omega}{B_{0}-1}\right).
\end{equation}
Before we continue it is informative to look at this result in two limits. The first is the adiabatic limit in which $\alpha\equiv \omega/(B_0 -1)  \ll 1 $. Here, $\alpha$ denotes the ratio between the system's relaxation timescale and that of the modulation. In the limit $\alpha\ll 1$ where the modulation is slow, the correction to the action reduces to $\Delta S_{SSR}= -\epsilon(B_0-1)/B_0$, which coincides with our adiabatic approximation result for the SSR, presented in subsection \ref{Adiabatic}. A second limit is the high frequency limit in which $\alpha \gg 1$, namely, when the modulation is fast. In this limit, however, the LT correction to the action becomes exponentially small in $\alpha$, and the dominant term in $\Delta S$ becomes the ${\cal O}(\epsilon^2)$ term~\cite{assaf2008population}, see Sec. \ref{Kapitsa}.

\begin{figure}[t]
	\includegraphics[width=0.85\linewidth]{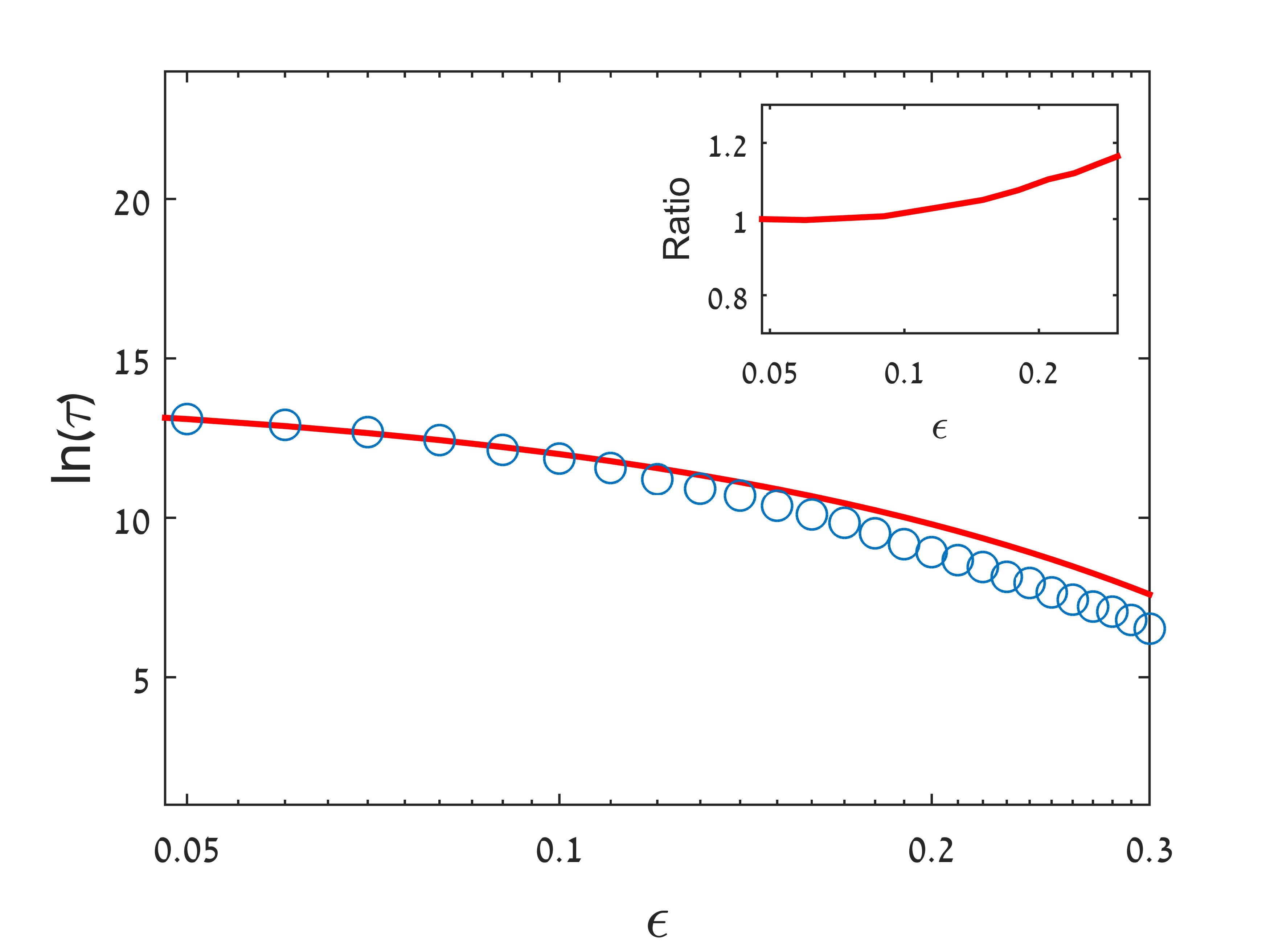}
	\caption{The logarithm of the MTE in the LT regime for the case of binomial (BN) SSD as a function of $\epsilon$: a comparison between the theoretical result (solid line) and Monte-Carlo simulations with time-dependent rates (symbols). The parameters are $B = 1.2$, $N = 3200$, $\omega = 0.24$, $m = 15 $, and $\rho = 0.4 $. Here the theoretical MTE (that does not include a preexponential prefactor) is multiplied by a constant so that it coincides with the numerical result at $\epsilon=0.05$. Inset shows the ratio between the theoretical and numerical results. Note that the range of $\epsilon$ is such that $N|\Delta S|\gtrsim {\cal O}(1)$, see text.}
	\label{fig:LT_B1_2N3200morePoints}
\end{figure}

In Fig. \ref{fig:ActionCalculationLT} we compare the theoretical action in the SSR case to numerical solutions of the Hamilton equations. This numerical solution also allows finding $t_0$ for which the action is minimal. In Fig. \ref{fig:LT_B1_2N3200morePoints} we compare the theoretical MTE [Eq.~(\ref{MTE})] in the case of a binomial SSD with numerical Monte Carlo simulations. The parameters for the binomial SSD are the number of trials $m$ and the probability of success in each trial $\rho$. In both figures the theoretical result holds as long as $\epsilon\ll 1$. A detailed description of the numerical solutions is found in Sec. \ref{NumericalCalc}.

\subsubsection{Linear theory - Bifurcation Limit} \label{LinearCorrectionintheBifurcationLimit}
For general SSDs, an analytical solution for the action in the LT regime can only be found close to the bifurcation limit, where $0<B_0-1\ll 1$. To this end we a-priori assume the momentum is small throughout the instanton trajectory (to be justified a-posteriori). We denote
$p_0 = (B_0-1)/[1+f^{\prime}(1)]\tilde{p}_0$, where $\tilde{p}_0={\cal O}(1)$, and $f^{\prime}(1)$ is found using L'Hôpital's rule
\begin{equation} \label{fprime}
f^{\prime}(1) = \frac{1}{2}\left(\frac{\sigma^2}{\langle k\rangle}+\langle k\rangle  -1\right).
\end{equation}
Substituting $p_0$ into Eqs. (\ref{problem}) and (\ref{Delta_S_simple}), keeping leading-order terms with respect to $B_0-1\ll 1 $, and minimizing the action with respect to $t_0$, we find that the minimum is obtained at $t_0 = \pi/\omega$. As a result, $\Delta S$ becomes
\begin{equation} \label{S_LT_Bifuraction}
\Delta S\approx-\epsilon\frac{\pi\omega}{1+f^{\prime}(1)}csch\left(\frac{\pi\omega}{B_0-1}\right),
\end{equation}
where the unpertubed action in this case satisfies~\cite{be2016rare}
\begin{equation} \label{S0_bifurcation}
S_{0}= \frac{1}{2}\frac{(B_0-1)^{2}}{1+f^{\prime}(1)}.
\end{equation}
Note, that in the SSR case where $f'(1)=0$, Eq.~(\ref{S_LT_Bifuraction}) reduces in the bifurcation limit to Eq.~(\ref{S_SSR}). Also note that, using  Hamilton equation~(\ref{Hamilton_q}) and (\ref{Hamilton_p}), the unperturbed instanton trajectory satisfies $q(t-t_0)=(B_0-1)/[1+e^{(B_0-1)(t-t_0)}]$ and $p(t-t_0)=-(B_0-1)/\{[1+f^{\prime}(1)][1+e^{-(B_0-1)(t-t_0)}]\}$, thus justifying a-posteriori our assumption regarding the smallness of the momentum. Finally, the result given by Eqs.~(\ref{S_LT_Bifuraction}) and (\ref{S0_bifurcation}) is valid as long as $S=S_0+\Delta S\gg 1/N$, which puts an upper limit on the value of $\epsilon$, depending on the value of $\alpha=\omega/(B_0 -1)$.

\subsection{Kapitsa Correction} \label{Kapitsa}

In this section we consider the high frequency limit, $\alpha\gg 1$, in which the modulation frequency is high compared to the typical relaxation rate of the system. The Kapitsa method was originally developed in the context of the ``Kapitsa pendulum'', see e.g. \cite{landau1976mechanics}, and here we apply a Hamiltonian extension of the method along the same lines of Ref.~\cite{assaf2008population}.

We begin with  Hamiltonian~(\ref{Hamiltonian}) and denote
\begin{equation} \label{Kapitsa_coordinates}
\begin{gathered}
q(t)=Q(t)+\frac{B_0}{B_0-1}\xi(t)\\p(t)=P(t)+\frac{B_0}{B_0-1}\eta(t)
\end{gathered}
\end{equation}
where $Q$ and $P$ are slowly-changing variables, and $\xi$ and $\eta$ are rapidly-changing, small corrections. Expanding $H(q,p,t)$ [given by  Eq.~(\ref{Hamiltonian})] around $q=Q$ and $p=P$ up to second order in $\xi$ and $\eta$ yields
\begin{eqnarray} \label{Kapitsa_expansion}
&&H(q,p,t)\approx H(Q,P,t)+\xi\frac{\partial H(Q,P,t)}{\partial Q} +\eta\frac{\partial H(Q,P,t)}{\partial P}\nonumber\\
&& +\xi^{2}\frac{\partial^{2}H(Q,P,t)}{\partial Q^{2}}  +\eta^{2}\frac{\partial^{2}H(Q,P,t)}{\partial P^{2}}+\xi\eta\frac{\partial^{2}H(Q,P,t)}{\partial P\partial Q}\nonumber\\
&&\equiv \tilde{H}(Q,P,t).
\end{eqnarray}
Using Eqs. (\ref{Kapitsa_coordinates}) and (\ref{Kapitsa_expansion}) the Hamilton equations become
\begin{eqnarray}
\dot{q}&=&\dot{Q}+\frac{B_0}{B_0-1}\dot{\xi}\simeq\frac{\partial\tilde{H}(Q,P,t)}{\partial P}\nonumber\\
\dot{p}&=&\dot{P}+\frac{B_0}{B_0-1}\dot{\eta}\simeq-\frac{\partial\tilde{H}(Q,P,t)}{\partial Q}.
\end{eqnarray}
Demanding that the rapidly oscillating terms balance each other, we find
\begin{eqnarray}
\dot{\xi}&=&\epsilon(B_0-1)\cos(\omega t)Q\left[(2P+1)f+P(P+1)f^{\prime}\right]\nonumber\\
\dot{\eta}&=&-\epsilon P(B_0-1)(P+1)f\cos(\omega t),
\end{eqnarray}
where $f=f(P+1)$. Treating $Q$ and $P$ as constants, we can solve these equations to find
\begin{eqnarray} \label{xi_and_eta}
\xi&=&\frac{\epsilon}{\alpha}\sin(\omega t)Q\left[(2P+1)f+P(P+1)f^{\prime}\right]\nonumber\\
\eta&=&-\frac{\epsilon}{\alpha}\sin(\omega t)P(P+1)f.
\end{eqnarray}
From this result it is clear that since $\alpha \gg 1$, $\epsilon\leq 1$ does not need to be small in order for this approximation scheme to be valid.

We now employ a canonical transformation to transform from the old $(q,p)$ to the new $(Q,P)$ variables, see Appendix A for details. The effective Hamiltonian, averaged over a period of a rapid oscillation $2 \pi /\omega$, becomes:
\begin{equation} \label{H_Kapitsa}
\bar{H}(Q,P) = H_0(Q,P) + \left(\frac{\epsilon}{\alpha}\right)^2 H_2(Q,P),
\end{equation}
where $H_0(Q,P)$ is the unperturbed Hamiltonian, given by Eq. (\ref{H_0}), and $H_2(Q,P)$ is given by Eq.~(\ref{H2}) in Appendix A. Since this effective Hamiltonian is time \textit{independent}, it is straightforward to find the effective instanton. Using Eq. (\ref{H_Kapitsa}) and repeating the steps that led to Eqs.~(\ref{Instanton}) and (\ref{Szero}), the instanton reads
\begin{equation} \label{Q_0_Kapitsa}
Q(P) = Q_0(P) + \left(\frac{\epsilon}{\alpha}\right)^2 Q_{K}(P)
\end{equation}
where $ Q_0(P)$ is the unperturbed instanton~(\ref{Instanton}), and $Q_{K}(P)$ is given by Eq.~(\ref{Q_0}) in Appendix A. As a result, the action becomes
\begin{equation} \label{SandSK}
S=-\int_{0}^{P_f}Q(P)dP=S_0 +\left(\frac{\epsilon}{\alpha}\right)^2 \Delta S_{K},
\end{equation}
where the second term, $\Delta S_K\equiv\int_{0}^{P_f}Q_{K}(P)dP$,  is the Kapitsa correction, while $S_0$ is given by Eq. (\ref{Szero}).

Let us demonstrate this method by explicitly calculating the Kapitsa correction for the SSR case. Here, $f(P)=1$, and $Q_K(P)$ given by Eq.~(\ref{Q_0}) becomes
\begin{equation}
Q_{K}(P)=\frac{1-B_0+3P-4B_0P-4B_0P^{2}}{2B_0}.
\end{equation}
Using the fact that in this case $P_f=-(B_0-1)/B_0$, the Kapitsa correction [Eq.~(\ref{SandSK})] becomes \cite{Bacaeer2015}
\begin{equation}
\Delta S_K=\frac{1}{4B_0}-\frac{1}{6}-\frac{1}{12B_0^{3}}.
\end{equation}

\subsubsection{Kapitsa correction - Bifurcation Limit}
We now briefly present the results of the Kapitsa correction close to the bifurcation $B_0-1\ll 1$, by repeating the steps done in Sec. \ref{LinearCorrectionintheBifurcationLimit}. Substituting $P=-(B_0-1)/[1+f^{\prime}(1)]\tilde{P}$ and $Q=\tilde{Q}(B_0-1)$ into $Q_K(P)$ [Eq.~(\ref{Q_0}) in Appendix A] and keeping only leading-order terms with respect to $B_0-1\ll 1$,  we have
\begin{equation}
\tilde{Q}(\tilde{P}) \approx \frac{1}{2}(\tilde{P}-1).
\end{equation}
Using this result and the fact that $P_f=-(B_0-1)/[1+f^{\prime}(1)]$ in this limit, Eq. (\ref{SandSK}) becomes
\begin{equation}
\Delta S_K = -\frac{1}{4}\frac{(B_0-1)^{2}}{1+f^{\prime}(1)}.
\end{equation}
Thus, the total action close to the bifurcation takes the following compact form
\begin{equation} \label{S0_Bifutcation_Kapitsa}
S = S_0\left[1-\frac{1}{2}\left(\frac{\epsilon}{\alpha}\right)^2\right],
\end{equation}
where $S_0$ is given by Eq. (\ref{S0_bifurcation}). Eq.~(\ref{S0_Bifutcation_Kapitsa}) is valid as long as $B_0-1\ll \omega\ll (B_0-1)^{-1/2}$, namely, the frequency cannot be too large. This is because on the one hand the Kapitsa method requires $\alpha\gg 1$ or $\omega\gg B_0-1$, while on the other hand, in $S$ we have  neglected ${\cal O}(B_0-1)^3$ terms, while keeping terms of ${\cal O}[(B_0-1)^2\alpha^{-2}]$ (here  $\epsilon\lesssim 1$).

\subsection{Adiabatic Approximation} \label{Adiabatic}
In the adiabatic limit the modulation frequency is much smaller than the typical relaxation rate of the system, i.e. $\alpha \ll 1$. In this case we can consider an approximation that is non-perturbative in the modulation amplitude. It has been shown by Assaf et el. \cite{assaf2008population} that the average extinction rate $\bar{r}_{ex}$ in the adiabatic limit is:
\begin{equation} \label{r_ex_average}
\bar{r}_{ex}=\frac{\omega}{2\pi}\int_{0}^{\frac{2\pi}{\omega}}r_{ex}(t^{\prime})dt^{\prime},
\end{equation}
with $r_{ex}(t)$ being the instantaneous value of the slowly time-dependant extinction rate. In this approximation the MTE is equal to $1/\bar{r}_{ex}$.

The mean extinction rate under bursty reproduction and constant reaction rates has been calculated by Be'er et al.~\cite{be2016rare}, including pre-exponential corrections. Following the steps outlined in the Appendix of Ref.~\cite{be2016rare}, the time-instantaneous extinction rate in our case is given by
\begin{equation} \label{r_ex_t}
r_{ex}(t)=A(t)e^{-NS[p_{f}(t), t]},
\end{equation}
with
\begin{equation} \label{adiabatic_action}
S(p, t)=\int_{p}^{0}\left[f(p^{\prime}\!+\!1)(1\!+\!\epsilon\cos(\omega t))-\frac{1}{B_0(p^{\prime}\!+\!1)}\right]dp^{\prime},
\end{equation}
and
\begin{eqnarray}
\hspace{-4.810mm}A(t) = -p_{f}\left[B_0(1\!+\!\epsilon\cos(\omega t))-1\right]\left[\frac{NS_{pp}(p_{f}, t)}{2\pi}\right]^{1/2}\!.
\end{eqnarray}
Here $p_f$ is explicitly time dependent and is defined by
\begin{equation} \label{p_f_adiabatic}
f(p_{f}+1)(1+\epsilon\cos(\omega t))=\frac{1}{B_{0}(p_{f}+1)},
\end{equation}
while $S_{pp}(p_{f}, t)$ is the second derivative of Eq. (\ref{adiabatic_action}) with respect to $p$ evaluated at $p=p_f(t)$. Substituting Eq. (\ref{r_ex_t}) into Eq.~(\ref{r_ex_average}), the average extinction rate is given by
\begin{equation}
\bar{r}_{ex}=\frac{\omega}{2\pi}\int_{0}^{\frac{2\pi}{\omega}}A(t^{\prime})e^{-NS[p_{f}(t), t^{\prime}]}dt^{\prime},
\end{equation}
which can be found via the saddle point approximation
\begin{equation}
\bar{r}_{ex}\approx \frac{\omega}{2\pi}A(t_s)e^{-NS[p_{f}(t_s), t_s]}\left[\frac{2\pi}{N \left|S_{tt}[p_f(t_s),t_s]\right|}\right]^{1/2}.
\end{equation}
Here the saddle point is found at $t_s = \pi/\omega$, and $S_{tt}[p_{f}(t_s), t_s]$ is the second derivative of action (\ref{adiabatic_action}) with respect to $t$ evaluated at $t_s$, while the time-dependent fluctuational momentum $p_f$ has to also be evaluated at $t_s$ according to Eq.~(\ref{p_f_adiabatic}). After some algebra, it can be shown that the average extinction rate becomes
\begin{equation} \label{adiabatic_rate_ex_final}
\bar{r}_{ex}\approx C e^{-NS[p_{f}(t_s),t_s]},
\end{equation}
where $S$ is given by Eq.~(\ref{adiabatic_action}). Here the pre-exponent
\begin{equation}
C = -p_{f}(B_0(1-\epsilon)-1)\left[\frac{(1\!-\!\epsilon)f^{\prime}(p_{f}\!+\!1)\!+\!\frac{1}{B_0(1+p_{f})^{2}}}{4\pi^2\epsilon\int_{p_{f}}^{0}f(p^{\prime}\!+\!1)dp^{\prime}}\right]^{1/2}\!,
\end{equation}
is independent on $N$ and the modulation frequency $\omega$, and $p_f$ is evaluated at $t_s$ according to Eq.~(\ref{p_f_adiabatic}).

\begin{figure}[t]
	\includegraphics[width=0.85\linewidth]{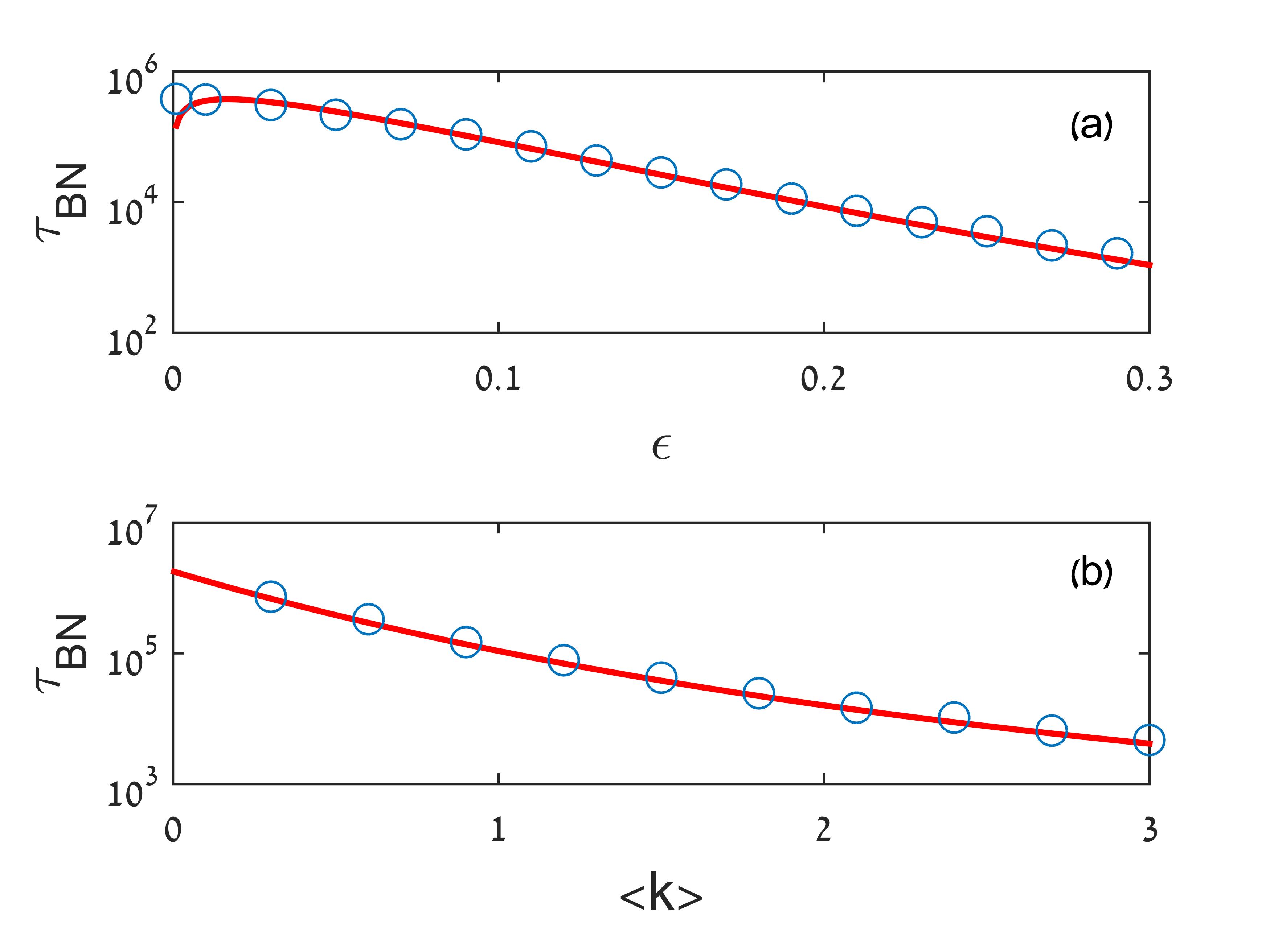}
	\caption{MTE for the case of binomial (BN) SSD in the adiabatic regime: theory (solid line) versus Monte Carlo simulations (symbols). In (a) the MTE is plotted against $\epsilon$, and the parameters are $B_0 = 3$, $\omega = 0.06$, $ m = 10 $, and $ \rho = 0.3 $. In (b) the MTE is plotted against the SSD's mean, $\langle k\rangle=m\rho$, and the parameters are $B_0 = 3$, $\omega = 0.06$, and $ \epsilon = 0.3 $.}
	\label{fig:AdiabaticMTEcombined}
\end{figure}

To illustrate this result, let us consider the SSR case for which $f(p)=1$. Here, Eq.~(\ref{adiabatic_rate_ex_final}) becomes
\begin{equation}
\bar{r}_{ex}=\left[\frac{(1-\epsilon)}{4\pi^2\epsilon}(B_{0}(1-\epsilon)-1)^3\right]^{1/2}e^{-N\left[S_0+\Delta S\right]},
\end{equation}
where $S_0 = 1-1/B_{0}-1/B_{0}\ln(B_0)$ in accordance with Eq.~(\ref{Szero}) for the SSR case, and $\Delta S = -1/B_{0}\ln(1-\epsilon)-\epsilon$. Note that for $\epsilon\ll 1$, this result reduces to $\Delta S = -\epsilon(1-1/B)$ in agreement with the LT result obtained in Sec. \ref{SSR_LT}. In Fig.~\ref{fig:AdiabaticMTEcombined} we compare theoretical results in the adiabatic limit for the case of a binomial SSD with Monte Carlo simulations and excellent agreement is observed.

The adiabatic theory is applicable as long as $\omega$ is much smaller, at all times, than the system's instantaneous relaxation rate, $B_0[1+\epsilon\cos(\omega t)] - 1$. This yields
\begin{equation} \label{ada_condition}
B_0(1-\epsilon) - 1\gg \omega,
\end{equation}
which also entails that $B_0(1-\epsilon) > 1$. That is, $\epsilon$ cannot be too close to $1$, otherwise the adiabatic approximation breaks down. In addition, we must have $S[p_f(t_s),t_s]\gg N^{-1}$ for the eikonal approximation to be valid. Finally, for the saddle-point approximation to be valid, the width around the saddle, $\left|S_{tt}[p_f(t_s),t_s]\right|^{-1/2}$, has to be much smaller than $\pi/\omega$, the distance between the saddle point and the integration boundaries in Eq. (\ref{r_ex_average}).

\section{Catastrophe} \label{Catastrophe}
Having considered time-periodic reaction rates, we now turn to the case of a catastrophe, which we model by a temporary drop in the population's birth rate. Here, the quantity of interest is not the MTE but rather the change in the extinction risk due to the catastrophe. Indeed, if the population dwells in a long-lived metastable state prior to extinction, before the catastrophe occurs the slowly time-dependent extinction probability satisfies  $\mathcal{P}_0(t) \equiv 1-e^{-t/\tau}$, where $\tau$ is the MTE of the population~\cite{assaf2010extinction,assaf2017wkb}. The catastrophe brings about an increase in the extinction risk $\Delta \mathcal{P}_0 $ due to the temporary decrease in the birth rate, and it is our goal in this section to calculate this change. Here, we generalize the treatment in Ref.~\cite{assaf2009population} which included the SSR case, and calculate the growth in the extinction risk for a general SSD.

\begin{figure}[t]
	\includegraphics[width=0.85\linewidth]{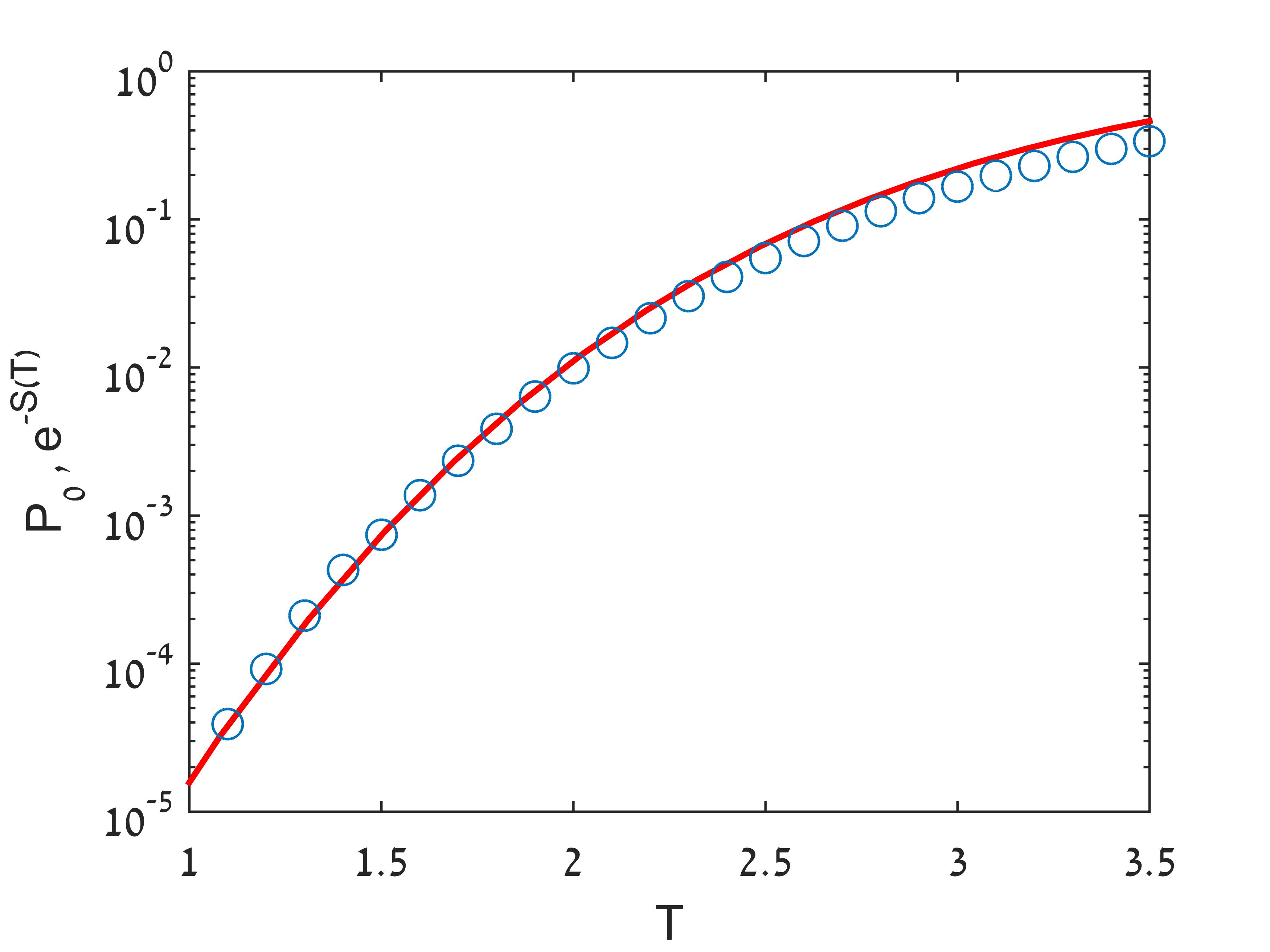}
	\caption{Probability of Extinction for the binomial (BN) SSD in the event of a catastrophe: theory (solid line) and Monte Carlo simulations (symbols), as a function of the catastrophe duration $T$. The parameters are $B=1.1$, $N=12000 $, $m= 10$ and $\rho= 0.3$. The theoretical result is multiplied by a constant prefactor to match the simulation result at $T=1.1$.}
	\label{fig:CatastropheN12000}
\end{figure}

To represent a catastrophe we substitute
\begin{equation}
g(t)=\begin{cases}
1 & t<0\;\;\text{or}\;\;t>T\\
0 & 0<t<T
\end{cases}
\end{equation}
into Hamiltonian (\ref{Hamiltonian}). The duration of the catastrophe is denoted by $T$, and we set it to start at some arbitrary time $t_{0}=0$. To proceed we calculate the different segments of the Hamiltonian, before, after and during the catastrophe, and then demand continuity between the different instanton solutions. The Hamiltonian before and after the catastrophe is the time-independent Hamiltonian [Eq.~(\ref{H_0})]. Whereas, during the catastrophe the birth rate vanishes and the Hamiltonian becomes
\begin{equation} \label{HamiltonianCata}
H_{c}(p,q)=-pq\left[1+B_{0}(p+1)q\right],
\end{equation}
which is independent on the specific choice of SSD. Now, in order to find the instanton it is necessary to match the instaton during the catastrophe to the pre- and post-catastrophe instanton. The instanton before and after is the zero energy line of Eq.~(\ref{H_0}) given by Eq.~(\ref{Instanton}). During the catastrophe, however, the a-priori unknown energy, $E_{c}=H_{c}$, is no longer zero, and is found by matching the non-zero energy line during the catastrophe
\begin{equation} \label{q_c}
q_{c}=\frac{1}{2B_{0}(p+1)}\left[\sqrt{1-\frac{4B_{0}(p+1)E_{c}}{p}}-1\right]
\end{equation}
with $q_{0}$. Solving $q_{0}=q_{c}$ gives us the intersections points $p_{1}(E_c)$ and $p_{2}(E_c)$
\begin{equation}
1+\sqrt{1-\frac{4B_{0}(p_{1,2}+1)E_{c}}{p_{1,2}}}=2B_{0}(p_{1,2}+1)f(p_{1,2}+1),
\end{equation}
which can be explicitly found for any particular choice of SSD. In order to determine $E_{c}$ we demand that the duration of the catastrophe be $T$. Putting $B(t)=0$ in Hamilton equation (\ref{Hamilton_p}) evaluated at $q=q_c$, using Eq.~(\ref{q_c}), and integrating from $t_0=0$ to $t=T$, we obtain
\begin{equation}\label{E_c}
T=\int_{p_{1}(E_{c})}^{p_{2}(E_{c})}\frac{dp}{\sqrt{p^{2}-p4B_{0}(p+1)E_{c}}},
\end{equation}
whose solution yields the energy $E_{c}$ associated with the catastrophe. Having found $E_c$, the action is given by~\cite{assaf2009population}
\begin{equation} \label{S_T}
S(T)=S_{0}-\int_{p_{2}(E_{c})}^{p_{1}(E_{c})}\left\{ q_{0}(p)-q_{c}\right\} dp-E_{c}T.
\end{equation}
According to the eikonal theory, this result for the decrease in action, together with Eq.~(\ref{E_c}), allows finding the increase in the extinction risk of the population up to exponential accuracy:
\begin{equation}\label{DelP}
\Delta \mathcal{P}_0 \sim e^{-N S(T)}.
\end{equation}
Note that this result is valid as long as $NS(T) \gg 1$. Also note that if $\Delta \mathcal{P}_0\gg \mathcal{P}_0$ (that is, if the catastrophe significantly increases the extinction risk), then Eq.(\ref{DelP}) approximately describes the extinction risk in the aftermath of the catastrophe. In Fig.~\ref{fig:CatastropheN12000} we compare the Eq.~(\ref{DelP}) with Monte Carlo simulations for the case of a binomial SSD. As expected, the theory holds as long as the duration $T$ is not too long such that $NS(T)\gg 1$.

While we have given a general recipe to find the increase in the population's extinction risk for a generic SSD, it is informative to examine these results close to the bifurcation limit where $B_0-1\ll 1$. In appendix \ref{Appendix2} we show that in this limit the analytical solution drastically simplifies, and the action can be written as
\begin{equation} \label{S_Bifurcation_Cata}
S(T)=\frac{2S_{0}}{e^{T}+1},
\end{equation}
with $S_0$ given by Eq.~(\ref{S0_bifurcation}). This result is a generalization of the result obtained by Assaf et el.~\cite{assaf2009population} for the case of the SSR, corresponding to $f(p)=1$.

\section{Numerical Calculations} \label{NumericalCalc}

\begin{figure}[t]
	\includegraphics[width=0.85\linewidth]{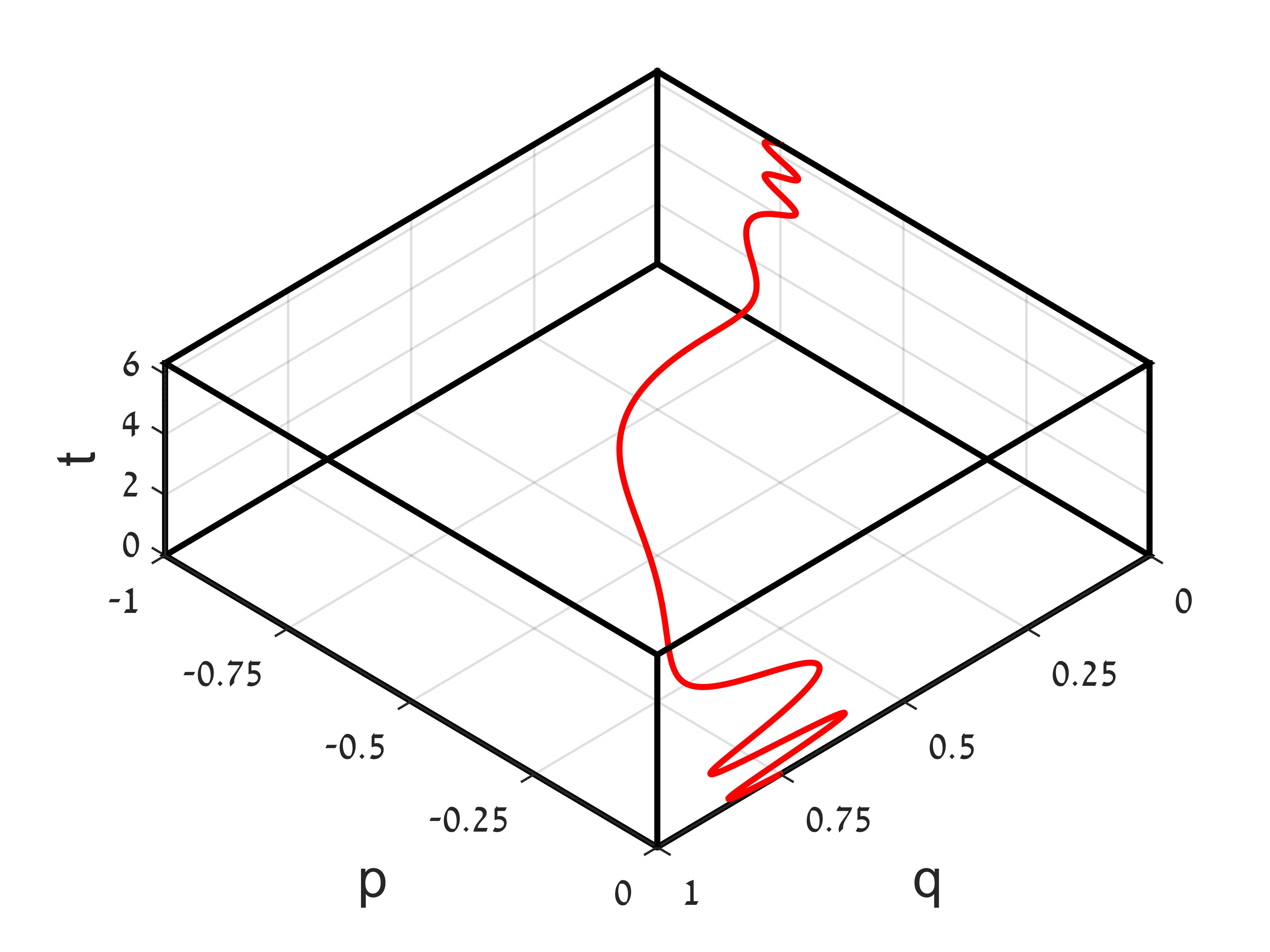}
	\caption{An example of an instanton trajectory of the perturbed Hamiltonian in the SSR case. The parameters are $B = 4$, $\epsilon = 0.3 $, and $\omega = 6$. The trajectory first performs large amplitude
oscillations around the mean-field fixed point $(q,p)=(0.75,0)$ and finally enters the vicinity of the fluctuational point $(q,p)=(0,-0.75)$.}
	\label{fig:phase_space_example}
\end{figure}

To verify our analytical results we have used two different numerical methods. The first method is a time-dependent Monte Carlo simulation. It is based on an extended version of the time-independent Gillespie algorithm~\cite{gillespie1977exact, gillespie1976general}, which accounts for time-dependent reactions rates, see \textit{e.g.}, Refs.~\cite{jansen1995monte, anderson2007modified}. In short, Gillespie's algorithm is composed of two steps: (i) advancing the time until the next reaction and (ii) choosing a reaction from all possible reactions, and updating the population size accordingly. The second step is insensitive to whether the reaction rates are explicitly time dependent, whereas accounting for bursty reproduction was done by considering all possible birth processes as independent reactions. To account for the time-dependent rates, we denote by $a_{\alpha\beta} $ the transition probability per unit time from state $\beta$ to state $\alpha$, and by $a_{\alpha}=\sum_{\beta}a_{\alpha\beta}$, the transition probability to reach $\alpha$ from all other states. At any given time $t$, the probability $P_{\alpha}$ that the system is still in configuration $\alpha$ after time $\delta t$ has elapsed is thus~\cite{jansen1995monte, anderson2007modified}
\begin{equation}
P_{\alpha}(t) = \exp\left[-\int_{t}^{t+\delta t}dt'a_{\alpha}(t')\right].
\end{equation}
In order to choose the time period $\delta t$ in which the next reaction will occur, we generate a random number from a uniform distribution in the interval $[0,1]$ and demand that this random number be equal to $P_{\alpha}$.  In the time-independent case, $\delta t$ can be explicitly found from this equation~\cite{gillespie1977exact}, but for time-dependent rates, this yields a transcendental equation \cite{anderson2007modified}, which has to be solved for each time step. Having found the time step $\delta t$, the next reaction is chosen according to the original Gillespie step, with the reactions rates evaluated at time $t+\delta t$~\cite{jansen1995monte}.

\begin{figure}[t]
	\includegraphics[width=0.85\linewidth]{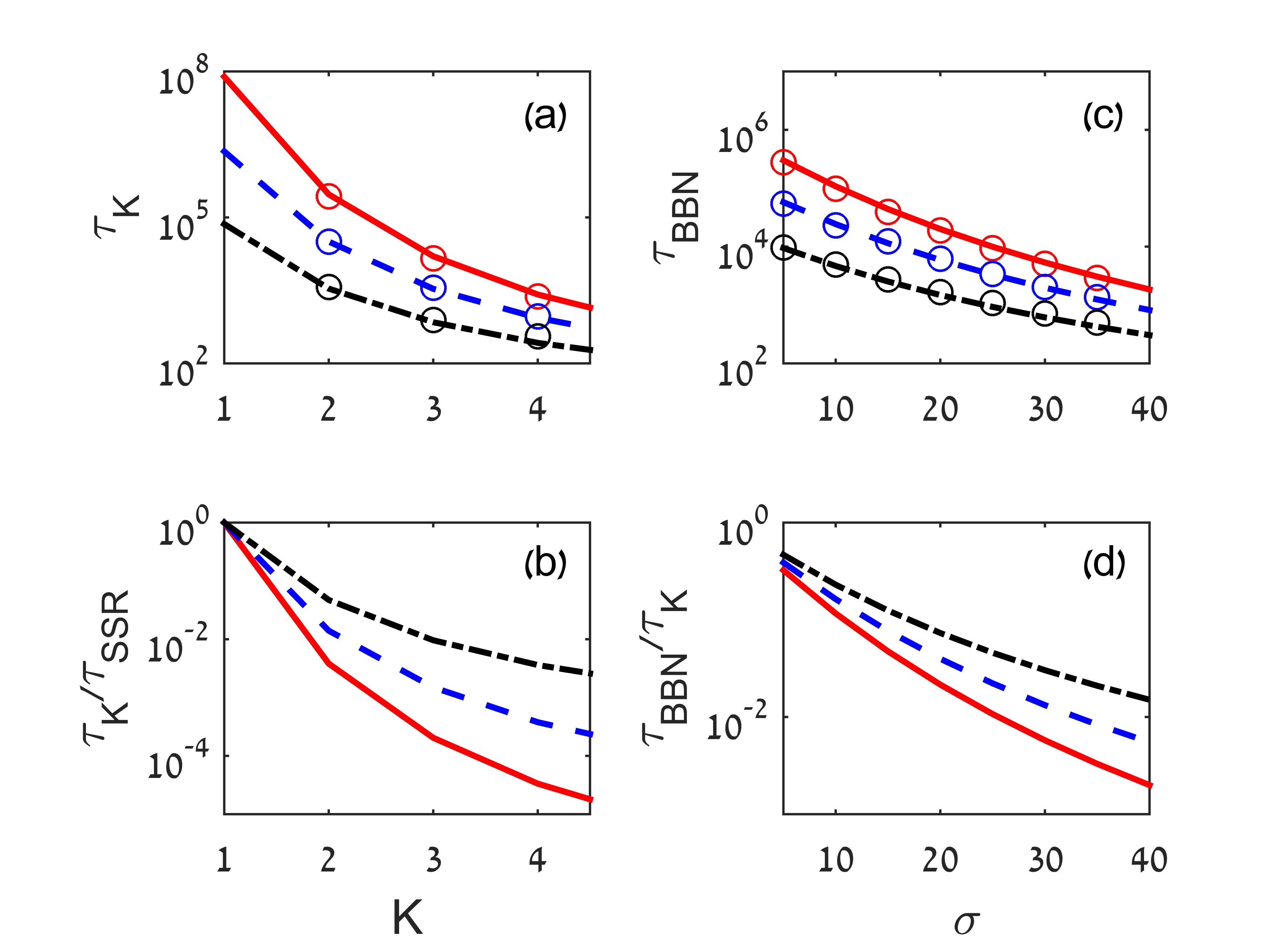}
	\caption{In (a) shown is the MTE for the case of K-step reaction, see text, as a function of $K$. Here the different lines correspond to the theoretical result in the adiabatic limit for $\epsilon = 0.05, 0.15, 0.25$ (solid, dashed and dash-dotted lines respectively), while the symbols are results of Monte-Carlo simulations. In (b) shown is the MTE in (a) normalized by the MTE in the case of the SSR, namely $K=1$. Parameters in (a) and (b) are $B = 3$, $N = 70$, and $\omega=0.01$. In (c) shown is the MTE for the case of BBN distribution, see text, as a function of the standard deviation $\sigma$, where the SSD's parameters ($m, \alpha, \beta$) are chosen to maintain a constant mean of $K=5$ (see text). The different lines correspond to the theoretical result in the adiabatic limit for $\epsilon = 0.05, 0.15, 0.25$ (solid, dashed and dash-dotted lines respectively), while the symbols are results of Monte-Carlo simulations. In (d) shown is the MTE in (c) normalized by the MTE in the case of the K-step reaction with $K=5$ and $\sigma=0$. Parameters in (c) and (d) are $B = 6$, $N = 80$, and $\omega=0.01$.}
	\label{fig:EffectsMeanStdCombined}
\end{figure}

When the MTE is long, employing such an algorithm, which includes solving a transcendental equation at each time step, may be extremely time consuming. As a result, we have also devised a numerical method to solve the explicitly time-dependent Hamilton equations numerically. For a time-independent Hamiltonian, finding the instanton numerically can be done directly using the shooting method. Here, we start at time $t=0$ in the close vicinity of the mean-field fixed point $(q_{mf}+\delta q, \delta p)$, where $\delta q,\delta p\ll 1$. To find the unstable eigendirection of the instanton, along which it leaves the vicinity of the mean-field fixed point at $t=0$, and enters at some final time, the close vicinity of the fluctuational point $(0,p_f)$, we substitute $q = q_{mf} +\delta q$ and $p = \delta p$ into the unperturbed instanton [Eq. (\ref{Instanton})]. Retaining leading-order terms, we arrive at $\delta q = [f'(1) +1/B]\delta p$, which determines the desired eigendirection. Having found the numerical solution to Eqs. (\ref{Hamilton_q}) and (\ref{Hamilton_p}) for some initial condition in the close vicinity of $(q_{mf},0)$ on the unperturbed instanton, one can find the action according to Eq. (\ref{Szero}).

In the time-dependent case, however, the initial conditions are more intricate to find. Here we start from the same initial conditions as for the time-independent case, but we now pay attention to the relative phase between the unperturbed instanton and the perturbed trajectory. As a result, we choose such relative phase as to minimize the action of the perturbed instanton. This relative phase in the numerical solutions is easily correlated to the minimization of the LT in subsection \ref{LT_section}, as both represent the deviation of the corrected trajectory from the original time-independent trajectory (see Fig. \ref{fig:ActionCalculationLT}).

\section{Summary and Discussion} \label{summary}
In this paper we have investigated a stochastic population under the joint influence of two non-demographic effects: a time-varying environment that gives rise to time-dependent reaction rates, and bursty reproduction that gives rise to uncertainty in the reaction step size. Two time-modulation protocols have been considered: a periodically-varying birth rate and a sudden temporary drop of the birth rate to zero. By using various analytical tools as well as extensive numerical simulations we have shown that such time modulation always decreases the MTE compared to the time-independent case. As a result, a time-varying environment always increases the extinction risk of a population. By accounting for bursty reprodcution with an arbitrary step-size distribution (SSD), this work generalizes previous works in this field which have treated constant-step-size reactions such as the Verhulst, or the branching-annihilation models.

How does bursty reproduction affect the extinction risk in the presence of time-dependent rates? In the time-independent case it has been shown by Be'er and Assaf that bursty reproduction increases the extinction risk compared to the SSR (single-step birth reaction)~\cite{be2016rare}. However, when compared with a birth reaction that produces exactly $K$ individuals (K-step reaction), depending on the skewness of the SSD, it has been shown that bursty reproduction can also decrease the extinction risk of the population~\cite{be2017enhancing}. Here we generalize these results by considering time-dependent rates and using a beta-binomial (BBN) distribution, which is a generalized version of the binomial distribution, see below.

\begin{figure}[t]
	\includegraphics[width=0.85\linewidth]{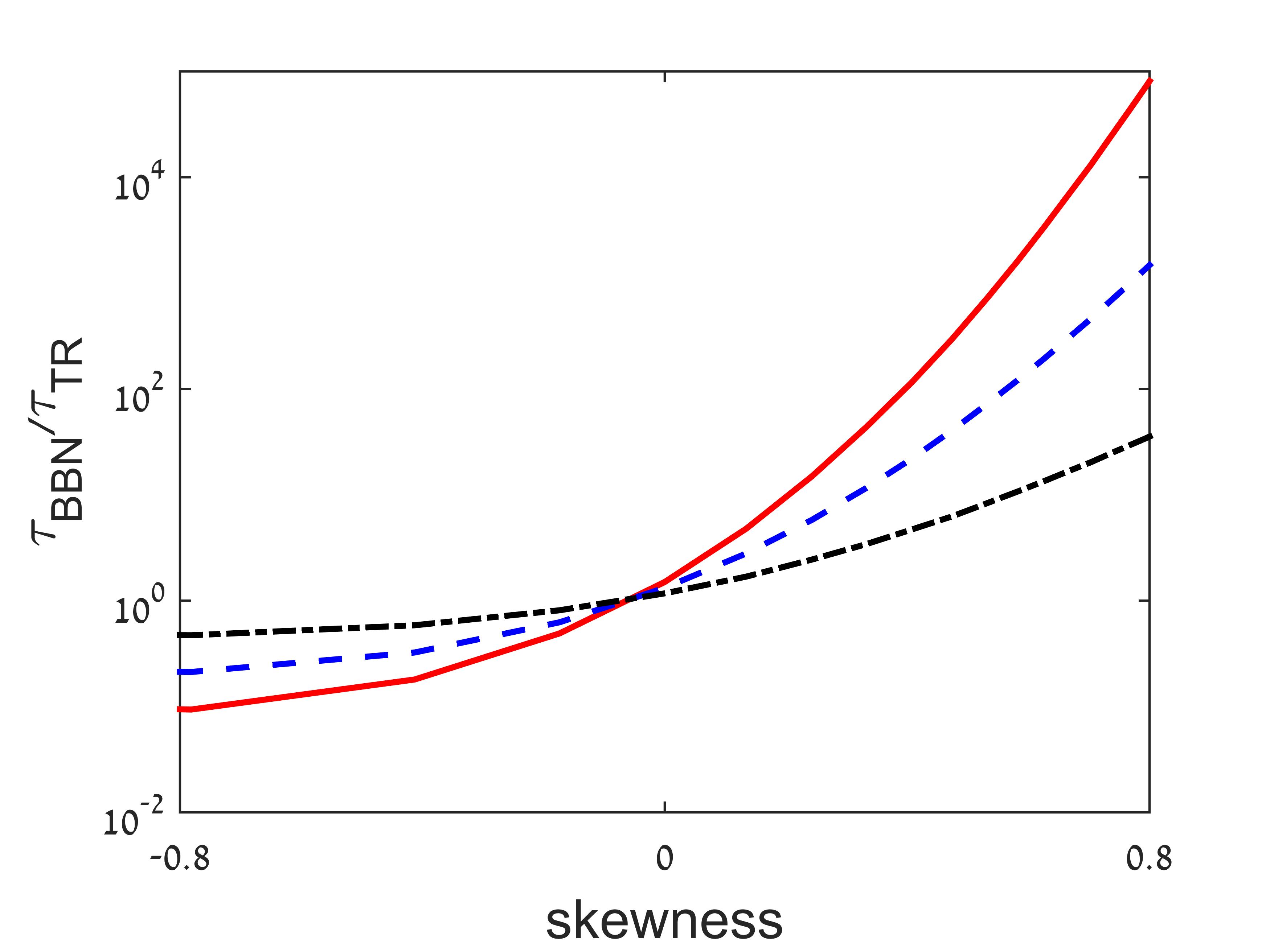}
	\caption{The ratio between the MTE in the BBN case and symmetric three value TR case (see text), as function of the BBN's skewness. The lines correspond to theoretical results in the adiabatic limit for $\epsilon = 0.05, 0.25, 0.45$ (respectively, solid, dashed and dash-dotted). The parameters are $B = 6$, $N=1000$, $\omega=0.01$, and SSD parameters are chosen such that the SSD's mean and variance be equal at each point.}
	\label{fig:skewnessTheoriticalFinal}
\end{figure}

In Fig. \ref{fig:EffectsMeanStdCombined} we study the dependence of the extinction risk on the first two moments of the SSD. In Fig. \ref{fig:EffectsMeanStdCombined}(a,b) we show that the MTE is exponentially reduced as the mean of the SSD is increased, by comparing the K-step reaction results with those using SSR. The reason for this increase in the population's extinction risk is that as the SSD's mean increases, birth events become less frequent and it is more likely to observe a series of death events that leads to population extinction. Yet, looking at the \textit{ratio} of the MTEs, this effect is significantly reduced when the rates are time dependent, see Fig. \ref{fig:EffectsMeanStdCombined}(b).

In Fig. \ref{fig:EffectsMeanStdCombined}(c,d) we study how the width of the SSD affects the extinction risk, by comparing the results of the K-step reaction with those using the BBN distribution. The latter is defined by three parameters: the number of independent trials $m$, and $\alpha,\beta$ which are the parameters of the beta distribution from which the probability of success of a single trial is taken. By tuning the parameters such that the mean of the BBN coincides with $K$, we show that the MTE is exponentially decreased when the SSD's width is increased, see Fig. \ref{fig:EffectsMeanStdCombined}(c). The reason for this increase in the population's extinction risk is that as the SSD's width increases, large-burst-size birth events become more likely and thus, it is more likely to observe a series of death events that drives the population to extinction. Yet, looking at the \textit{ratio} of the MTEs, again the effect of MTE reduction is drastically reduced when introducing time-dependent rates, see Fig. \ref{fig:EffectsMeanStdCombined}(d).

We have also examined how the SSD's third moment affects the population's extinction risk. In Fig. \ref{fig:skewnessTheoriticalFinal} we compare the BBN results with those of a symmetric three-value triangular (TR) SSD. To study the net effect of the third moment, the latter is tuned such that the mean and variance coincide with that of the BBN distribution. Figure \ref{fig:skewnessTheoriticalFinal}  demonstrates that when the SSD is positively skewed, the MTE is increased and vice versa, while for zero skewness the MTEs almost coincide. This is because for positively-skewed SSDs (here the BBN), the median is smaller than the mean, and thus, small-burst-size birth events are more likely than in the TR case, where the median equals the mean. Yet, similarly to the cases of the first and second moments, as the amplitude of the time modulation $\epsilon$ increases, the effect of increasing the MTE as the skewness increases, is diminished~\footnote{Close to the bifurcation, this effect vanishes since the MTE depends only on the first two moments of the SSD.}.

To understand the interplay between having time-dependent rates and bursty reproduction, we look at the adiabatic limit (Sec. \ref{Adiabatic}). Here, the system ``waits" until the effective birth rate goes to its minimum, $B_0(1-\epsilon)$ (see Sec. \ref{Adiabatic}), and only then it goes to extinction. As a result, the typical population size, which directly depends on the birth rate, and from which the system goes extinct, is decreased. Therefore, since the MTE is exponentially sensitive to the typical population size, we find that the effect of increase/decrease in the extinction risk is exponentially diminished due to the time-dependent rates. Finally, note that while Figs.~\ref{fig:EffectsMeanStdCombined} and \ref{fig:skewnessTheoriticalFinal} demonstrate the adiabatic regime, we have checked that this effect (although weaker) still exists in the non-adiabatic regime.

\appendix

\section{Kapitsa Results} \label{Appendix1}
In this appendix we provide some intermediate results for the high frequency limit, $\alpha=\omega/(B_0-1) \gg 1$. Using Eqs. (\ref{Kapitsa_coordinates})  and (\ref{xi_and_eta}) we perform an almost canonical
transformation from the old variables $q$ and $p$ to the new variables $Q$ and $P$:

\begin{equation} \label{p_KapitsaA}
p(Q,P,t)=P\left[1-\frac{\epsilon}{\alpha}(P+1)f\sin(\omega t)\right]
\end{equation}

\begin{multline} \label{q_KapitsaA}
q(Q,P,t) =\frac{Q}{1-\frac{\epsilon}{\alpha}\sin\left(\omega t\right)\left[\left(2P+1\right)f+P\left(P+1\right)f^{\prime}\right]}\\ \approx Q\left\{1+\frac{\epsilon}{\alpha}\sin(\omega t)\left[(2P+1)f+P(P+1)f^{\prime}\right]\right. \\ \left.+\left(\frac{\epsilon}{\alpha}\right)^2\sin^{2}(\omega t)\left[(2P+1)f+P(P+1)f^{\prime}\right]^{2}\right\}.
\end{multline}
This transformation is canonical up to third order in $\mathbb{\mathcal{O}}[(1/\alpha)^{3}]\ll 1$, since the Poisson brackets satisfy $ \left\{ q,p\right\} _{Q,P}= 1+\mathbb{\mathcal{O}}[(1/\alpha)^{3}]$. The generating function of this transformation satisfies~\cite{landau1976mechanics}
\begin{equation} \label{canonical_transformation}
F_{2}(q,P,t)=qP\left[1-\frac{\epsilon}{\alpha} (P+1)f\sin(\omega t)\right].
\end{equation}
This allows making the transformation $H^{\prime}=H+\partial F_{2}/\partial t$, where by time-averaging the new Hamiltonian $H^{\prime}$ over a period of rapid oscillation $2 \pi /\omega$, we arrive at the effective time-independent Hamiltonian~(\ref{H_Kapitsa}). Here the correction to the unperturbed Hamiltonian, due to the high-frequency time modulation, satisfies:
\begin{eqnarray} \label{H2}
&&H_2 = \frac{1}{2}QP\left[B_0P(1+P)^{2}f^{2}-P(1+P)\right.\nonumber\\
&&(1+3P+4B_0Q(1+P)(1+2P))f f^{\prime}\nonumber\\
&&-P^{2}(1+P)^{2}(1+3B_0Q(1+P))(f^{\prime})^{2}\nonumber\\
&&+f^{2} \left(-\left(P+2P^{2} +B_0Q(1+P)(1+5P(1+P))\right) \right.\nonumber\\
&&+\left.\left. B_0P(1\!+\!P)^{2}\left((1\!+\!2P)f^{\prime}+\frac{1}{2}P(1\!+\!P)f^{\prime\prime}\right)\right)\right].
\end{eqnarray}
Finally, this correction brings about a correction to the unperturbed instanton~(\ref{Instanton}), which has the form
\begin{eqnarray} \label{Q_0}
&& Q_{K}(P) = \frac{1}{4B_{0}}\left[-2B_0(1\!+\!2P)^{2}f^{3}+4P^{2}(1\!+\!P)(f^{\prime})^{2}\right.\nonumber\\
&&+2Pff^{\prime}\left(3+5P -3B_{0}P(1+P)^{2}f^{\prime}\right)+ f^{2}\left(2+6P \right.\nonumber\\
&&\left.\left. +B_{0}P(1+P)\left(-6(1+2P)f^{\prime}+P(1+P)f^{\prime\prime}\right)\right)\right].
\end{eqnarray}
This result allows to explicitly calculate the correction to the action in Eq. (\ref{SandSK}), see Sec.~\ref{Kapitsa} in the main text.

\section{Catastrophe Calculations in the Bifurcation Limit} \label{Appendix2}
In this appendix we calculate the action in the case of a catastrophe close to the bifurcation limit. The treatment here goes along the same lines as in Sec.~\ref{LinearCorrectionintheBifurcationLimit}. We denote $p_0 = (B_0-1)/[1+f^{\prime}(1)\tilde{p}_0$ and $q_0 = (B_0-1)\tilde{q}_0$, where $\tilde{p}_0$ and $\tilde{q}_0$ are ${\cal O}(1)$. We also denote $H=\tilde{H}(B_{0}-1)^{2}/[f^{\prime}(1)+1]$. Therefore, in the leading order the Hamiltonian before and after the catastrophe reduces to
\begin{equation}
\tilde{H}=\tilde{p}\tilde{q}(\tilde{p}-\tilde{q}+1)(B_{0}-1),
\end{equation}
while the normalized instanton is $\tilde{q}=1+\tilde{p}$.
The Hamiltonian during the catastrophe [Eq.~(\ref{HamiltonianCata})] reduces in leading order to
\begin{equation}
\tilde{H}_{c}=-\tilde{p}\tilde{q}.
\end{equation}
Demanding that $\tilde{E}_{c} = \tilde{H}_{c}$, the non-zero energy trajectory during the catastrophe becomes $\tilde{q}_{c}=-\tilde{E}_{c}/\tilde{p}$. The intersection points between the instantons before/after and during the catastrophe are found by solving $\tilde{q}_{c}=\tilde{q}$:
\begin{equation}
\tilde{p}_{1,2}=-\frac{1}{2}(1\pm\sqrt{1-4\tilde{E}_{c}}).
\end{equation}
During the catastrophe the Hamilton equation for the momentum reduces to $\dot{\tilde{p}}=\tilde{p}$, which yields $\tilde{p}_{2}/\tilde{p}_{1}=e^{T}$. As a result we find~\cite{assaf2009population}
\begin{eqnarray}
\hspace{-5mm}&&\tilde{E}_{c}=\frac{e^{T}}{(1+e^{T})^{2}}=\frac{1}{4}\cosh^{-2}(T/2),\\
\hspace{-5mm}&&\tilde{p}_{1}=-\frac{1}{2}[1\!+\!\tanh(T/2)]\,,\;\;\tilde{p}_{2}=-\frac{1}{2}[1\!-\!\tanh(T/2)].
\end{eqnarray}
Finally, using Eq.~(\ref{S_T}) we arrive at
\begin{equation}
S(T)=\frac{(B_0-1)^2}{2[1+f^{\prime}(1)]}\left[1-\tanh(T/2)\right]=\frac{2S_{0}}{e^{T}+1},
\end{equation}
where we have used the definition of $S_0$ from Eq.~(\ref{S0_bifurcation}).


\bibliographystyle{apsrev4-1}
\bibliography{VA_bib}

\end{document}